\documentclass[12pt,fullpage,fleqn,psfig]{article}
\usepackage{latexsym}
\usepackage{epsfig}

\large
\newcommand{\beq}{\begin{equation}}
\newcommand{\eeq}{\end{equation}}
\newcommand{\ba}{\begin{array}{ccc}}
\newcommand{\ea}{\end{array}}
\newcommand{\nn}{\nonumber}

\newcommand{\be}{\begin{eqnarray}}
\newcommand{\ee}{\end{eqnarray}}

\begin{document}
\newcommand{\mat}{\left ( \begin{array}{cc}}
\newcommand{\emat}{\end{array} \right )}
\newcommand{\matt}{\left ( \begin{array}{ccc}}
\newcommand{\ematt}{\end{array} \right )}
\newcommand{\matf}{\left ( \begin{array}{cccc}}
\newcommand{\ematf}{\end{array} right )}
\newcommand{\vect}{\left ( \begin{array}{c}}
\newcommand{\evect}{\end{array} \right )}
\newcommand{\Tr} {\rm Tr}
\newcommand{\cotanh}{{\rm cotanh}}
\def\beqn{\begin{eqnarray}}
\def\eeqn{\end{eqnarray}}
\def\la{\lambda}
\def\ga{\gamma}
\def\om{\omega}
\def\al{\alpha}
\def\d{\partial}
\def\Tr{ {\rm Tr} }

\renewcommand{\baselinestretch}{2}

\thispagestyle{empty}
\parskip=4mm
\newcommand{\R}{{\mathchoice{\hbox{$\sf\textstyle I\hspace{-.15em}R$}}
{\hbox{$\sf\textstyle I\hspace{-.15em}R$}} {\hbox{$\sf\scriptstyle
I\hspace{-.10em}R$}} {\hbox{$\sf\scriptscriptstyle
I\hspace{-.11em}R$}}}}

\renewcommand{\theequation}{\arabic{section}.\arabic{equation}}

\hfill{SUNY-NTG-01/40}
%\begin{flushright}
%SUNY-NTG-01-99
%\end{flushright}

\vspace{1cm}
\begin{center} 
{\Large\bf Diquark Condensate in QCD with Two Colors  \\\vspace{.5cm}
  at Next-to-Leading  Order}
\vspace{8mm}

K. Splittorff$^1$, D. Toublan$^2$ and J.J.M. Verbaarschot$^3$

{\small
$^1$NBI, Belgdamsvej 17, Copenhagen, Denmark \\
$^2$Physics Department, University of Illinois at Urbana-Champaign,
Urbana, IL 61801, USA\\  
$^3$Department of Physics and Astronomy, SUNY, Stony Brook,
NY\,11794, USA\\} 
\vskip 1cm

{\bf Abstract}

\end{center}
We study QCD with two colors and quarks in the fundamental representation
at finite baryon density in the limit of light quark masses. 
In this limit the free energy of this theory reduces to the free 
energy of a chiral Lagrangian which is 
%determined by 
based on the symmetries
of the microscopic theory. In earlier work this Lagrangian was 
analyzed at the mean field level and a phase transition to 
a phase of condensed diquarks was found at a chemical potential of
half the diquark mass (which is equal to the pion mass). In this article
we analyze this theory at next-to-leading order in chiral perturbation
theory. We show that the theory is renormalizable and calculate the
next-to-leading order free energy in both phases of the theory.
By deriving a  Landau-Ginzburg theory for the order parameter we show
that the 
finite one-loop contribution and the next-to-leading order terms
in the chiral Lagrangian do not qualitatively change the phase transition.
In particular, the critical chemical potential is equal to half the
next-to-leading order pion mass, and the phase transition is second order.  


\setcounter{page}{0}
\newpage
\setcounter{equation}{0}
\section{Introduction}

QCD with two colors and quarks in the fundamental
representation of the gauge group is in many respects similar to 
QCD with three colors and quarks in the fundamental representation.
Both theories are confining and asymptotically free, both
exhibit a spontaneous  breaking of chiral symmetry, and both are believed
to be in a different phase at high temperature and density.
However, the
mechanism of the phase transition at nonzero baryon chemical potential
has to be different. The nucleon in QCD with three colors is a
fermion with mass much larger than the pion mass. At finite baryon density
a phase transition may occur if the Fermi-surface becomes unstable. 
Instead, for two colors,
the lightest baryon is a boson with the same mass as the usual pionic
Goldstone bosons. 
It is a diquark state. 
For a quark-chemical potential equal to half the mass of the
lightest baryon a phase 
transition to a Bose-Einstein condensate of diquarks takes place
very much like the
transition to a pion condensate for increasing pion chemical potential in ordinary
QCD. This phase transition can be
described completely in terms of a low-energy effective theory 
for the
Goldstone modes associated with the spontaneous breaking of 
chiral symmetry \cite{KST,KSTVZ}.
This low-energy effective theory essentially relies on
the symmetries of the underlying microscopic theory. 
In earlier work \cite{KST,KSTVZ}
this theory was studied at the mean field level. A phase transition 
from the normal phase to
a phase of condensed diquarks was found at a chemical potential equal
to  half the pion mass. This phase is a superfluid phase with massless
Goldstone bosons. 
One of them  results from the spontaneous breaking of the
$U(1)$ group associated with baryon number.

For QCD with three colors and fundamental quarks, the fermion determinant
at nonzero chemical potential has a complex phase making first principle
Monte Carlo simulations impossible. On the other hand,
for QCD with two colors the fermion determinant remains real at nonzero
chemical potential and Monte Carlo simulations can be performed for an
even number of flavors
\cite{SU2lattice.old,SU2lattice.new}. 
We are thus in a unique situation where both analytical
results and Monte-Carlo simulations are available in the nonperturbative
domain of QCD at nonzero baryon density. The hope is that this ultimately
will lead to a better understanding of dense quark matter in QCD with
three colors and fundamental quarks. What already has been achieved is 
that the mean field results have been confirmed
by  lattice QCD simulations of several independent groups
\cite{LATTICE}.

The same analysis can be made in several other cases. They can be
classified  
according to the Dyson index of the Dirac operator of the underlying
microscopic theory \cite{HOV,KSTVZ}. 
For QCD with two colors and $N_f$ quarks in the fundamental
representation the Dyson index is $\beta =1$
with Goldstone manifold given by $SU(2N_f)/Sp(2N_f)$. 
For QCD with two or more colors and quarks in the adjoint representation
the Dyson index is $\beta=4$ and the Goldstone manifold is given by
$SU(2N_f)/O(2N_f)$. 
The third case with Dyson index $\beta =2$ is QCD with
three or more colors and quarks in the fundamental representation. 
Although chiral perturbation theory
is irrelevant for full QCD at nonzero baryon chemical potential, 
the low-energy effective theory for $\beta=2$ 
can be applied to phase-quenched QCD \cite{TV}, i.e. QCD 
at nonzero chemical potential with the fermion
determinant replaced by its absolute value. Phase quenched QCD has
been reinterpreted as QCD at finite isospin density \cite{Misha-Son} 
and both theories are described by the same low-energy effective theory. 
This theory
has been generalized to include chemical potentials for the different
quark species \cite{SSS,Dominique}. The theories
for $\beta=1$, $\beta =2$ and $\beta=4$ display a 
similar phase diagram at finite
baryon chemical potential. They can be studied within very similar
effective field theories \cite{KSTVZ}. It is also possible to write
down effective theories for  the Goldstone sector  of
microscopic theories with an imaginary vector potential but without
an involutive automorphism such as the axial symmetry of the Dirac
operator.  The main difference is the structure of the Goldstone 
manifold which is determined by the pattern of spontaneous symmetry
breaking.
Such theories  enter 
in the context of disordered condensed matter systems  \cite{Fyodorov,Efetov}
and the effective
theory and its phase diagram are similar to that of QCD-like theories.
They are also relevant to
QCD in 3 dimensions at nonzero chemical potential \cite{gernot}.
As is the case in chiral
perturbation theory at zero
chemical potential \cite{OTV,DOTV,TV,tiloch,Kim-Poul,taka,guhr,Poul} 
these mean field
results can also be obtained from a chiral 
Random Matrix Model \cite{VJ} at nonzero chemical potential.

Up to now the above effective theories have only been analyzed at the mean
field level \cite{KST,KSTVZ,TV,Misha-Son,SSS,Dominique}. 
In this work we perform a one-loop calculation for the case
$\beta =1$. Our first goal is to show that the theory is
renormalizable in both phases,
i.e. that for an arbitrary background field
the one-loop divergences can be absorbed into the coupling constants
of the next order Lagrangian. Our second goal is to investigate the one-loop
corrections to the mean field results for the phase transition. 
Our expectation is that the critical chemical potential of the 
next-to-leading order theory is equal to half the renormalized 
pion mass at this order. Related questions such as the effect 
of one-loop corrections on the order of the phase transition and 
the critical exponents will be addressed as well.

The paper is organized as follows. In section 2 we give a brief review of
the low-energy effective theory for QCD with two colors and nonzero
chemical 
potential. The next-to-leading order effective Lagrangian is introduced
in section 3. The one-loop corrections of the leading order effective
Lagrangian are calculated for an arbitrary background field and it is shown
that the divergences can be absorbed into the coupling constants of the
next-to-leading order effective Lagrangian.
In section 4 we calculate
the one-loop corrections to the free energy in the normal phase and the 
phase of condensed diquarks.  A Landau-Ginzburg model
for the phase transition is derived from the effective theory
 in section 5, and its parameters
are calculated from chiral perturbation theory. In section 6, we
discuss the  number density, the
chiral condensate and the diquark condensate. Concluding remarks are made
in section 7. Some of the technical details are worked out in two 
appendices. In Appendix A we calculate the next-to-leading order
corrections 
to the pion mass and the pion decay constant. In Appendix B the one-loop
integrals for the baryonic Goldstone modes in the diquark phase are
evaluated.

\section{Low-Energy Limit of QCD with Two Colors}

In this section we review the low-energy effective theory of QCD with
two colors and fundamental quarks. More details can be found in the
original literature \cite{V-Smilga,TV14,KST,KSTVZ}.

\subsection{The Lagrangian of QCD with Two Colors}
\label{sect:symQC2D}

The QCD partition function is given by the ensemble average 
determinant of the Dirac operator. For two colors the ensemble average is over
$SU(2)$-valued gauged fields weighted by the usual gluonic action.
The Dirac operator for quark mass $m$
and chemical potential $\mu$ is given by
\be
D= \gamma_\nu D_\nu +m +\mu\gamma_0.
\ee 
Here, $D_\nu \equiv \partial_\nu +A_\nu$ is the covariant derivative,
$\gamma_\mu$ are the Euclidean Dirac matrices, and $A_\nu$ are
the $SU(2)$-valued gauge fields.
Many of the differences between QCD with gauge group 
$SU(2)$ and fundamental quarks and
QCD with three or more colors and fundamental quarks originates from
the pseudo-reality of $SU(2)$. 
It
manifests itself through the
anti-unitary symmetry of the Dirac operator $D$ ($C$ is the charge
conjugation operator and $\tau_2$ acts in
color space) 
\be
D\tau_2C\gamma_5 = \tau_2C\gamma_5D^*.
\ee
The reality of the fermion 
determinant  is a consequence of this symmetry. However in our context
a more important consequence is that for $N_f$ fundamental quarks the
flavor symmetry group is enlarged to $SU(2N_f)$ which is sometimes
referred to as the Pauli-G\"ursey symmetry \cite{pauli-gursey}.
In a basis given by left-handed
quarks, $q_L$, and conjugate right-handed anti-quarks, $\sigma_2\tau_2
q_R^*$,  
($\sigma_2$ acts in flavor space) this enlarged symmetry group acts on the 
flavor indices of the spinors
\be
\Psi \equiv \left(\begin{array}{c} q_L \\ \sigma_2\tau_2 q_R^*
\end{array}\right) \ 
\qquad
{\rm as} \qquad
\Psi \to V\Psi, \; {\rm with} \; V \in SU(2 N_f).
\label{2Nf-rotation}
\ee
This symmetry becomes manifest by writing 
the Lagrangian for QCD with two colors in terms of these spinors:
\be
\label{Lqcd}
L_{\rm QCD} & = & i\Psi^\dagger \sigma_\nu(D_\nu-\mu B \delta_{0\nu})\Psi
-(\frac{1}{2}\Psi^T \sigma_2\tau_2 {\cal M} \Psi + {\rm h.c.})\ , % +, jac
\ee
where $\sigma_\nu=(-i,\sigma_k)$. The baryon charge matrix 
denoted by $B$ is given by
\be
 B\equiv\left( \begin{array}{cc}
      1&0\\0&-1 \end{array} \right ), 
\ee
and the matrix that includes the mass term and the diquark source term
is defined by \cite{KSTVZ}
\be
\label{chi}
  {\cal M}\equiv     \sqrt{m^2+j^2} 
  ({\hat M}  \cos  \phi+ {\hat J} \sin \phi),
\ee
where $\tan\phi = j/m$ and
\be
  {\hat M} & \equiv & \left( \begin{array}{cc} 0&1\\-1&0
    \end{array} \right) \  , \ \ \ 
      {\rm and} \ \ \ \hat J\equiv\left(
    \begin{array}{cc} 
      iI&0\\0& iI \end{array} \right)  \ \  {\rm with} \ \ I\equiv \left( \begin{array}{cc} 0&-1\\1&0
    \end{array} \right).
\label{MandB}
\ee
Here and in the rest of the paper we consider only even $N_f$.
% added a sentence - K
The matrices $\hat{M}$ and $\hat J$ correspond to the mass term and to
the diquark source, respectively.
It is clear that only the kinetic term has the full $SU(2N_f)$ 
symmetry whereas it is broken explicitly by the  chemical potential
and the source terms.

What is more important is the spontaneous breaking of chiral symmetry
by the formation of a chiral condensate. It breaks the flavor symmetry
in the same way as the mass terms, i.e. according to
$SU(2N_f) \to Sp(2N_f)$. 
Along with the chiral condensate
it is possible to form a diquark condensate. This condensate preserves the
same symmetries as its source term, i.e. $Sp(N_f)\times Sp(N_f)$. 
At nonzero chemical potential and mass the symmetry is $SU(N_f)\times
U_B(1)$. 
The diquark condensate  breaks this symmetry spontaneously according to 
$SU(N_f)\times U_B(1) \to Sp(N_f)$. The baryon number is no longer
a conserved quantity. 

Although the source terms and the chemical potential break the flavor
symmetry, the full flavor symmetry can be retained if the
source terms are transformed according to
\begin{eqnarray}
\label{globalQCDtrans}
 {\cal M}
  \rightarrow V^* {\cal M} V^\dagger, \hspace{1.0cm} 
 B \rightarrow V B V^\dagger \ .
\end{eqnarray}
This global $SU(2N_f)$ symmetry of the Lagrangian can be 
extended to a local flavor symmetry  by introducing the vector field
$B_\nu=(B, 0,0,0)$ with transformation properties
\cite{KST,KSTVZ,Alvarez} 
\begin{eqnarray}
\label{localQCDtrans}
   B_\nu \rightarrow V B_\nu V^\dagger -\frac1\mu V \partial_\nu
   V^\dagger. 
\end{eqnarray}

Because of the spontaneous breaking of chiral symmetry, the low-energy
limit of QCD with two colors is a theory of weakly interacting Goldstone
bosons. The Goldstone manifold is given by $SU(2N_f)/Sp(2N_f)$. If the
chiral condensate is denoted by $\bar{\Sigma}$ it can be parameterized by
\begin{eqnarray}
  \Sigma=U \bar{\Sigma} U^T,
\label{Sigma}
\end{eqnarray}
where
\begin{eqnarray}
  U=\exp (i \Pi/2 F) \hspace{1cm} {\rm and} \hspace{.5cm} \Pi=\pi_a 
  X_a/\sqrt{2 N_f}.
\end{eqnarray}
Here, $F$ is the pion decay constant,
the fields $\pi_a$ are the Goldstone modes and the $X_a$ are the
$2 N_f^2-N_f-1$ generators of the coset $SU(2N_f)/Sp(2N_f)$.
They obey the relation
$X_a \bar \Sigma
=\bar \Sigma X_a^T$ and are 
normalized according to ${\rm Tr} X_a X_b=2 N_f \delta_{ab}$.

The Lagrangian of the Goldstone modes is obtained by the requirement that
it has the same symmetry properties as the underlying microscopic theory.
The Lagrangian for $\Sigma$ should be invariant under the {\em local}
transformations
\be
\Sigma \to V \Sigma V^T, \qquad V \in SU(2N_f)
\ee
if the source terms and $B_\nu$ are transformed according to 
(\ref{globalQCDtrans},\ref{localQCDtrans}). We adopt the usual power
counting 
of chiral perturbation theory that 
the square root of the quark mass, the square root of the diquark source,
the chemical potential  and the momenta are of the same order. Then to
order $p^2$ 
it is simple to write down a Lagrangian
that is invariant under the global transformations (\ref{globalQCDtrans}). 
Invariance under the local transformation (\ref{localQCDtrans}) can be 
achieved by introducing the covariant derivative \cite{KST}
\begin{eqnarray}
  \nabla_\nu \Sigma&=&\partial_\nu \Sigma-\mu (B_\nu \Sigma+\Sigma 
  B_\nu^T), \nonumber \\
  \nabla_\nu \Sigma^\dagger&=&\partial_\nu \Sigma^\dagger+\mu (B_\nu^T
  \Sigma^\dagger+\Sigma^\dagger B_\nu).
\label{CovDeriva}
\end{eqnarray}
The effective Lagrangian to leading order in the momentum
expansion having the required invariance properties is 
thus given by \cite{KSTVZ}
\begin{eqnarray}
  \label{L2}
  {\cal L}^{(2)}=\frac{F^2}{2} {\rm Tr} \Big[ \nabla_\nu \Sigma
  \nabla_\nu \Sigma^\dagger - \chi^\dagger \Sigma - \chi
  \Sigma^\dagger \Big],
\end{eqnarray}
where we have introduced
the source term 
\be
\chi = \frac G{F^2} {\cal M}^\dagger.
\ee
To leading order in the chiral expansion
$G$ is related to the
quark-antiquark condensate at $j=\mu=T=0$ according to  
$\langle\bar \psi \psi\rangle_0=2 N_f G$, and 
the pion mass is given by the Gell-Mann--Oakes--Renner relation: 
$  M^2=m G/F^2$.

For zero diquark source, 
the leading-order Lagrangian (\ref{L2}) describes two different
phases: 
a normal phase and a phase of condensed diquarks. They are
separated by a second order phase transition at $\mu_c=M/2$.
The chiral condensate in both phases can be parameterized as
\cite{KSTVZ}
\begin{eqnarray}
  \label{min}
  \bar{\Sigma}(\alpha)=\cos \alpha \Sigma_c+ \sin \alpha \Sigma_d,
\end{eqnarray}
where
\begin{eqnarray}
\begin{array}{lll}
\alpha=0 & {\rm if} & \mu<\mu_c \\
\cos\al=\frac{M^2}{4\mu^2} & {\rm if} & \mu>\mu_c,
\end{array}
\end{eqnarray}
and
\begin{eqnarray}
  \Sigma_c\equiv I \ \ \  {\rm and}
\ \ \ \Sigma_d\equiv \left( \begin{array}{cc} iI & 0 \\ 0 & iI
  \end{array} \right). 
\end{eqnarray}
A non-zero value of $\al$ corresponds to diquark condensation
\cite{KSTVZ}. 
The orientation of the condensate (\ref{min}) minimizes the
static part of the leading-order effective Lagrangian (\ref{L2}).
At nonzero $\alpha$ it is sometimes useful to introduce rotated 
generators
of $SU(2 N_f)/Sp(2 N_f)$ defined by
\be
X_a(\al) = V_\al X_a(\al=0) V_\al^\dagger \qquad {\rm with}\qquad
V_\al=\exp(-\frac{\al}{2}\Sigma_d\Sigma_c).
\ee

At nonzero $j$ 
the diquark condensate is always non-vanishing.
In this case, the saddle point is still given by
$\bar \Sigma=\cos \alpha \Sigma_c+\sin \alpha \Sigma_d$,
but $\alpha$ is determined by the saddle point equation
\begin{eqnarray}
  \label{SadPt}
  4 \mu^2 \cos \alpha \sin \alpha=\tilde M^2 \sin (\alpha-\phi),
\end{eqnarray}
where $\tilde M^2=G \sqrt{m^2+j^2}/F^2$.  
Also in this case it may be useful to introduce rotated generators.

The spectrum contains $2 N_f^2-N_f-1$ pseudo-Goldstone bosons. In the 
normal phase, $\alpha=0$, they can be distinguished by their baryon
charge. There are $N_f^2-1$ ``usual'' pions 
with charge zero, denoted by $P$, $N_f(N_f-1)/2$ diquarks with charge
$+2$, denoted by $Q$, and $N_f(N_f-1)/2$ antidiquarks with charge
$-2$, 
denoted by $Q^\dagger$. The inverse of their propagator is
respectively
given by \cite{KSTVZ},
\begin{eqnarray}
  \label{invPropN}
  D^P&=&p^2+M^2, \nonumber \\
  D^Q&=&\left( \begin{array}{cc} p^2+M^2-4 \mu^2 & 4 i \mu p_0 \\ 
                4 i \mu p_0 & p^2+M^2-4 \mu^2 
\end{array} \right).
\end{eqnarray}

In the phase of condensed diquarks the pseudo-Goldstone modes of the
normal 
phase are mixed into four different types.
We now have $N_f (N_f+1)/2$ $P_S$-modes,
$(N_f^2-N_f-2)/2$ $P_A$-modes   
and $N_f (N_f-1)$ $Q$-modes. 
The inverse of their respective propagators reads \cite{KSTVZ}
\begin{eqnarray}
  \label{invPropD}
  D^{P_S}&=&p^2+M_1^2 +\frac 14 M_3^2,
 \nonumber \\
D^{P_A}&=&p^2+M_2^2 +\frac 14 M_3^2,
\\
  D^Q&=&\left( \begin{array}{cc} p^2+M_1^2 & i M_3 p_0 \\ 
                i M_3 p_0 & p^2+M_2^2
\end{array} \right), \nonumber
\end{eqnarray}
where 
\begin{eqnarray}
\label{massQ}
M_1^2&=&\tilde M^2 \cos (\alpha-\phi)-4 \mu^2 \cos 2\alpha \nonumber
\\ 
M_2^2&=&\tilde M^2 \cos (\alpha-\phi)-4 \mu^2 \cos^2\alpha \\
M_3^2&=&16 \mu^2 \cos^2 \alpha. \nonumber
\end{eqnarray}
By diagonalizing the inverse propagator for the $Q$-modes we find
$N_f (N_f-1)/2$ $\tilde Q$-modes
and  $N_f (N_f-1)/2$ $\tilde Q^\dagger$-modes. The $\tilde Q$-modes
are the true massless Goldstone modes of the superfluid phase at $j=0$
and $\mu>M/2$ \cite{KSTVZ}.

\section{ Next-to-Leading Order Effective Theory}

We wish to examine the effect of all one-loop diagrams to the
free energy.
Such contributions are of $O(p^4)$ in the power counting of Chiral 
Perturbation Theory, that is next-to-leading order in the momentum 
expansion \cite{GaL}. In this section we shall construct and
renormalize 
Chiral Perturbation Theory to $O(p^4)$ for QCD with two colors and 
fundamental quarks. We will  closely follow the work
of Gasser and Leutwyler for QCD with three colors and fundamental 
quarks \cite{GaL}.

At next-to-leading order in chiral perturbation theory, 
the partition function
contains four different contributions: the tree graphs of the
$O(p^2)$ effective Lagrangian ${\cal L}^{(2)}$, the one-loop
diagrams from ${\cal  L}^{(2)}$, the tree graphs of
the $O(p^4)$ effective Lagrangian ${\cal L}^{(4)}$ and the
contribution from the axial anomaly \cite{GaL}. The free energy
at next-to-leading order can thus be written as
\begin{eqnarray}
  \label{NLOpartfct}
  S=S_2+S_{1\rm{-loop}}+S_4+S_A,
\end{eqnarray}
where $S_2=\int dx {\cal L}^{(2)}$, $S_{1\rm{-loop}}$ is the
one-loop contribution to the free energy from the  one-loop diagrams of
${\cal L}^{(2)}$, $S_4=\int dx {\cal L}^{(4)}$, and $S_A$ is the
Wess-Zumino-Witten functional which reproduces the axial anomaly.

\subsection{Operators at Next-to-Leading Order}

For QCD with three colors and $N_f$
quarks in the fundamental representation
the most general effective Lagrangian to $O(p^4)$ has already been
constructed in the literature  \cite{GaL,NfChPT}. The
  $O(p^4)$ effective Lagrangian for QCD with two
colors and $N_f$ quarks in the fundamental representation can be found
along the same lines. Although
the Goldstone manifold is different in the two
cases, the same operators appear in the Lagrangian. The $O(p^4)$
effective Lagrangian contains all the Lorentz
invariant operators of $O(p^4)$ that are invariant
under local $SU(2 N_f)$ flavor transformations.
Keeping only the terms that
are relevant to  our case, we deduce from \cite{GaL,NfChPT} the effective
Lagrangian
\begin{eqnarray}
  \label{L4}
  {\cal L}^{(4)}&=&-L_0 {\rm Tr} \Big[ \nabla_\nu \Sigma \nabla_\tau
  \Sigma^\dagger  \nabla_\nu \Sigma \nabla_\tau
  \Sigma^\dagger \Big] - L_1 \Big({\rm Tr} \Big[  \nabla_\nu \Sigma
\nabla_\nu  \Sigma^\dagger   \Big]\Big)^2 \nonumber \\
&&- L_2 {\rm Tr} \Big[  \nabla_\nu \Sigma \nabla_\tau
  \Sigma^\dagger   \Big] {\rm Tr} \Big[  \nabla_\nu \Sigma \nabla_\tau
  \Sigma^\dagger   \Big] -L_3 {\rm Tr} \Big[  \Big(\nabla_\nu \Sigma
 \nabla_\nu  \Sigma^\dagger \Big)^2  \Big] \\
&&+L_4 {\rm Tr} \Big[ \chi \Sigma^\dagger
  + \Sigma \chi^\dagger \Big] {\rm Tr} \Big[  \nabla_\nu \Sigma
  \nabla_\nu 
  \Sigma^\dagger   \Big]+ L_5 {\rm Tr} \Big[ \Big(\chi \Sigma^\dagger
  + \Sigma \chi^\dagger \Big) \Big( \nabla_\nu \Sigma \nabla_\nu
  \Sigma^\dagger   \Big)  \Big]  \nn \\
&&- L_6 \Big({\rm Tr} \Big[ \chi \Sigma^\dagger
  + \Sigma \chi^\dagger \Big]\Big)^2-L_7 \Big( {\rm Tr} \Big[
  \chi^\dagger 
  \Sigma-\chi \Sigma^\dagger \Big] \Big)^2
  \nonumber \\
&&-L_8 {\rm Tr} \Big[ \chi \Sigma^\dagger
  \chi \Sigma^\dagger+\Sigma \chi^\dagger \Sigma \chi^\dagger
  \Big]-H_2 {\rm Tr} \Big[ \chi \chi^\dagger  \Big]. \nonumber
\end{eqnarray}
The low-energy coupling constants $L_i$, $i=1,\dots,8$, and $H_2$ are
'bare' coupling constants.
The term related to the coupling constant $H_2$ is a contact term
that, in our case, will enter only in the vacuum energy and in the
quark-antiquark condensate. Because we only consider quarks with equal
masses, the coupling constant $L_7$ will not
appear in any of our results.

\subsection{Renormalization to One Loop}

In order to compute the one-loop contributions from ${\cal L}^{(2)}$,
we expand ${\cal L}^{(2)}$ in the fluctuations around the solution of
the classical equation of motion following the
analysis of Gasser and Leutwyler \cite{GaL}. The one-loop
contributions from ${\cal L}^{(2)}$ are UV divergent. They have to be
regularized in some way, and the regularized divergences must be
canceled by  a renormalization of the coupling constants that appear
in ${\cal L}^{(4)}$. If the theory is renormalizable the values
of 
the counter-terms must be independent of $\mu$, $M$, and $j$, and,
especially, the 
renormalization constants must be identical in the two phases. We show
that this is indeed the case.

The solution of the classical equation of motion in the normal phase
can be 
written as
\begin{eqnarray}
  \label{class}
  \tilde{\Sigma}=U_\alpha \bar \Sigma U_\alpha^T,
\end{eqnarray}
where $U_\alpha$ belongs to the Goldstone manifold $SU(2
N_f)/Sp(2N_f)$  
with $Sp(2N_f)$ subgroup that leaves $\bar \Sigma$ invariant. 
Using that the $\bar \Sigma$ can be viewed as a
rotation of $\Sigma_c=I$, i.e. $\bar \Sigma = V_\alpha \Sigma_c
V_\alpha^T$ with $V=\exp(-\alpha \Sigma_d \Sigma_c/2)$,
we can also parameterize the Goldstone manifold as follows
\begin{eqnarray}
  \tilde \Sigma=V_\alpha U V_\alpha^\dagger \bar \Sigma
  V_\alpha^* U^T V_\alpha^T=V_\alpha U I U^T V_\alpha^T=V_\alpha U^2 I
  V_\alpha^T.
\end{eqnarray}
The last equality holds because $U=U_{\al=0}$.
The expansion around $\tilde{\Sigma}$ is obtained by the substitution
\be 
U \to U e^{i\xi},
\ee
where $\xi$ is a linear combination
of the generators $X_a$ of the Goldstone manifold $SU(2N_f)/Sp(2 N_f)$
satisfying $X_a I =I X_a^T$,
\begin{eqnarray}
\label{xiCompon}
\xi=\sum_{a=1}^{2  N_f^2-N_f-1} \xi^a X^a,
\end{eqnarray}
In other words, 
$\xi$ is a  $2 N_f \times 2 N_f$ hermitian traceless matrix that
satisfy $\xi I=I \xi^T$. The expansion about $\tilde \Sigma$ can thus
can be written as
\begin{eqnarray}
  {\Sigma}&=&V_\alpha U \left(1+ i \left(\frac {\xi}2 \right) -\frac12
  \left( \frac {\xi}2 \right)^2 +\cdots \right) I
  \left( 1+i \left(\frac {\xi^T}2 \right)-\frac12 \left( \frac{\xi^T}2
    \right)^2 +\cdots \right) U^T V_\alpha^T \nonumber \\
&=&V_\alpha U \left( 1+ i \xi-\frac12 \xi^2 +\cdots \right) U I
V_\alpha^T.
\end{eqnarray}

The $O(p^2)$ effective Lagrangian is given by (\ref{L2}).
Therefore, to second order in the fluctuations, we find
\begin{eqnarray}
  \label{L2exp}
  {\cal L}^{(2)}=& &\frac{F^2}2 {\rm Tr}[\nabla_\nu \tilde \Sigma
  \nabla_\nu 
  \tilde \Sigma^\dagger-\chi \tilde \Sigma^\dagger
-\tilde \Sigma  \chi^\dagger] \nonumber \\
&+&\frac{F^2}2 {\rm Tr}[\nabla_\nu^\alpha ( U \xi U )
\nabla_\nu^\alpha 
  (U^\dagger \xi U^\dagger )-\frac12 \nabla_\nu^\alpha ( U^2 ) 
\nabla_\nu^\alpha  (U^\dagger \xi^2 U^\dagger) \\
&-&\frac12 \nabla_\nu^\alpha (U^\dagger)^2
  \nabla_\nu^\alpha (U \xi^2 U)+\frac12 \xi^2 
(U I \chi^{\alpha \dagger} U-U^\dagger  \chi^\alpha I U^\dagger)],
\nonumber 
\end{eqnarray}
where the superscript $\alpha$ indicates that the original sources
have been replaced by the rotated sources:
\begin{eqnarray}
  \chi^\alpha=V_\alpha^\dagger \chi V_\alpha^* \hspace{2cm} {\rm and}
  \hspace{2cm} B_\nu^\alpha=V^\dagger_\alpha B_\nu V_\alpha,
\end{eqnarray}
and the covariant derivative $\nabla_\nu^\alpha$ is  given by
$\nabla_\nu$ defined in (\ref{CovDeriva})with $B_\nu$ replaced by
$B_\nu^\alpha$. Notice that $B_\nu^\alpha=I B_\nu^\alpha I$.

Let us denote the classical action by $\tilde S_2$. 
Up to second order in the fluctuations the action can be written as
\begin{eqnarray}
  \int dx {\cal L}^{(2)}=\tilde S_2+\frac{F^2}2 \int dx {\rm
    Tr}\Big({\rm 
    d}_\nu \xi {\rm d}_\nu \xi- [\Delta_\nu,\xi]
  [\Delta_\nu,\xi]-\sigma \xi^2 \Big),
\end{eqnarray}
where
\begin{eqnarray}
\label{def}
  {\rm d}_\nu \xi&=&\partial_\nu \xi + [\Gamma_\nu, \xi] \nonumber \\
  \Gamma_\nu&=&\frac12 U^\dagger \partial_\nu U-\frac12 U\partial_\nu
  U^\dagger-\frac\mu2 U^\dagger B_\nu^\alpha U-\frac\mu2 U
  B_\nu^{\alpha T}
  U^\dagger \nonumber  \\
  \Delta_\nu&=&\frac12 U^\dagger \nabla_\nu(U^2) U^\dagger=-\frac12 U
  \nabla_\nu (U^\dagger)^2 U \\
  \sigma&=&\frac12 (U^\dagger \chi^\alpha I U^\dagger-U I \chi^{\alpha
    \dagger} U).
 \nonumber
\end{eqnarray}
Notice that with the above conventions $\nabla_\nu(U \xi U)=U ({\rm d}_\nu
\xi+\{\Delta_\nu,\xi\})U$, and that the field strength associated with
$\Gamma_\nu$ reduces to $\Gamma_{\mu \nu}=\partial_\mu
\Gamma_\nu-\partial_\nu
\Gamma_\mu+[\Gamma_\mu,\Gamma_\nu]=-[\Delta_\mu,\Delta_\nu]$.
In terms of the components of $\xi=\sum_a \xi^a X^a$ where $X^a$ are
the generators of $SU(2 N_f)/Sp(2 N_f)$ with ${\rm Tr} X^a X^b=2 N_f
\delta^{ab}$, we obtain
\begin{eqnarray}
  \int dx {\cal L}^{(2)}=\tilde S_2-{N_f F^2} (\xi, D\xi),
\end{eqnarray}
with scalar product defined by $(f,g)=\sum_a \int dx f^a(x) g^a(x)$, 
and the
differential operator $D$ is given by
\begin{eqnarray}
  D^{ab}\xi^b&=&{\rm d}_\nu {\rm d}_\nu \xi^a+\hat{\sigma}^{ab} \xi^b
  \\ 
  {\rm d}_\nu \xi^a&=&\partial_\nu \xi^a+\hat{\Gamma}_\nu^{ab} \xi^b
  \\ 
  \hat{\Gamma}_\nu^{ab}&=&-\frac1{2N_f} {\rm Tr} \Big( [X^a,X^b]
  \Gamma_\nu   \Big)\\
  \hat{\sigma}^{ab}&=&\frac1{2N_f} {\rm Tr}\Big( [X^a,\Delta_\nu] \;
  [X^b,\Delta_\nu]  \Big) +\frac1{4N_f} {\rm Tr} \Big( \sigma
  \{X^a,X^b \}  \Big). 
\end{eqnarray}
Notice that the field strength associated with $\hat{\Gamma}_\mu$ is
simply given by $\hat{\Gamma}^{ab}_{\mu\nu}=-\frac12 {\rm Tr}(
[X^a,X^b] \Gamma_{\mu\nu})$.

The one-loop correction to the free energy 
is obtained by performing a Gaussian integral resulting in
\begin{eqnarray}
  S_{1{\rm-loop}}=\frac12 \ln {\rm det} D.
\end{eqnarray}
As is usual in ChPT \cite{GaL} we choose to regularize the determinant by means
dimensional regularization \cite{hooft-veltman}. There are ultraviolet
divergences that produce poles in $d$. The standard calculation leads
to \cite{GaL, NfChPT}
\begin{eqnarray}
  S_{1{\rm-loop}}=\sum_{a,b} \int d^d x
  \Big(\frac1d \delta^{ab}
  -\frac1{4 \pi (d-2)} \hat{\sigma}^{ab} \delta^{ab}  + \frac1{(4
    \pi)^2 (d-4)} 
 \Big\{ \frac1{12}  \hat{\Gamma}_{\mu\nu}^{ab}
 \hat{\Gamma}_{\mu\nu}^{ba}
+\frac12  \hat{\sigma}^{ab} \hat{\sigma}^{ba} \Big\} + \cdots \Big)
\nonumber \\
\label{ZfromD}
\end{eqnarray}
In order to perform the sum over the $SU(2 N_f)/Sp(2 N_f)$
generators, we need the following two formulae
\begin{eqnarray}
\sum_{a=1}^{2 N_f^2-N_f-1} {\rm Tr}[X^a A] {\rm Tr}[X^a B] &=&
N_f {\rm Tr} [A B]-N_f {\rm Tr}[A I B^T I]- {\rm Tr}[A] {\rm
Tr}[B] \\
\sum_{a=1}^{2 N_f^2-N_f-1} {\rm Tr}[X^a A X^a B] &=&-
{\rm Tr} [A B]+N_f {\rm Tr}[A I B^T I]+N_f {\rm Tr}[A] {\rm Tr}[B].
\end{eqnarray}
~From these two formulae we easily derive the identities
\begin{eqnarray}
\sum_{a,b} \hat{\Gamma}_{\mu\nu}^{ab} \hat{\Gamma}_{\mu\nu}^{ba}
&=& 2 (N_f-1) {\rm Tr}[\Gamma_{\mu \nu} \Gamma_{\mu \nu}]\\
 &=& 4 (N_f-1) \Big( {\rm Tr}[\Delta_\mu \Delta_\nu \Delta_\mu
\Delta_\nu]-{\rm Tr}[(\Delta_\mu \Delta_\mu)^2] \Big),
\end{eqnarray}
and 
\begin{eqnarray}
\sum_{a,b} \hat{ \sigma}^{ab} \hat{\sigma}^{ba} &=&2 N_f{\rm
Tr}[(\Delta_\mu \Delta_\mu)^2]+ \Big( {\rm Tr}[\Delta_\mu
\Delta_\mu]\Big)^2 \nonumber +2 {\rm Tr}[\Delta_\mu \Delta_\nu]
{\rm Tr}[\Delta_\mu \Delta_\nu] \nonumber \\
&&-{\rm Tr}[\sigma] {\rm Tr}[\Delta_\mu \Delta_\mu]-2 N_f {\rm
Tr}[\sigma \Delta_\mu \Delta_\mu] \nonumber \\
&&+ \frac{N_f^2+1}{4 N_f^2} \Big( {\rm Tr}[\sigma] \Big)^2+
\frac{(N_f+1)(N_f-2)}{2 N_f} {\rm Tr}[\sigma^2],
\end{eqnarray}
where we have used that $X_a I=I X_a^T$, $I \Delta_\mu^T
I=-\Delta_\mu$, $I \sigma^T I=-\sigma$, and
$I \Gamma_{\mu \nu}^T I=-\Gamma_{\mu \nu}^\dagger
=\Gamma_{\mu \nu}$. Therefore using (\ref{def}) we find
\begin{eqnarray}\label{RENOtrace}
 \sum_{a,b} \Big\{ \frac1{12}  \hat{\Gamma}_{\mu\nu}^{ab}
\hat{\Gamma}_{\mu\nu}^{ba}+\frac12
  \hat{\sigma}^{ab} \hat{\sigma}^{ba} \Big\} 
  &=&\frac{N_f-1}{48} {\rm Tr} \Big[ \nabla_\mu \tilde 
\Sigma \nabla_\nu
  \tilde \Sigma^\dagger  \nabla_\mu \tilde \Sigma \nabla_\nu
  \tilde \Sigma^\dagger \Big] +\frac1{32} \Big({\rm Tr} \Big[
  \nabla_\mu \tilde \Sigma \nabla_\mu
  \tilde \Sigma^\dagger   \Big]\Big)^2 \nonumber \\
&&\hspace{-3cm}+\frac1{16} {\rm Tr} \Big[  \nabla_\mu \tilde
\Sigma \nabla_\nu
  \tilde \Sigma^\dagger   \Big] {\rm Tr} \Big[  \nabla_\mu 
\tilde \Sigma \nabla_\nu
  \tilde \Sigma^\dagger   \Big] +\frac{2 N_f+1}{48} {\rm Tr} \Big[
  \Big(\nabla_\mu \tilde \Sigma \nabla_\mu 
\tilde \Sigma^\dagger \Big)^2  \Big] \\
&&\hspace{-3cm}-\frac1{16} {\rm Tr} \Big[ \chi \tilde
\Sigma^\dagger
  + \tilde \Sigma \chi^\dagger \Big] 
{\rm Tr} \Big[  \nabla_\mu \tilde \Sigma \nabla_\mu
  \tilde \Sigma^\dagger   \Big]
- \frac{N_f}8 {\rm Tr} \Big[ \Big(\chi \tilde \Sigma^\dagger
  + \tilde \Sigma \chi^\dagger \Big) \Big( \nabla_\mu 
\tilde \Sigma \nabla_\mu
  \tilde \Sigma^\dagger   \Big)  \Big] \nonumber \\
&&\hspace{-3cm} +\frac{N_f^2+1}{32 N_f^2} \Big({\rm Tr} \Big[
\chi \tilde \Sigma^\dagger + \tilde \Sigma \chi^\dagger
\Big]\Big)^2
 +\frac{(N_f+1)(N_f-2)}{16 N_f} {\rm Tr} \Big[\chi
\tilde \Sigma^\dagger
  \chi \tilde \Sigma^\dagger+\tilde \Sigma \chi^\dagger 
\tilde \Sigma \chi^\dagger  +2 \chi \chi^\dagger  \Big]. \nonumber
\end{eqnarray}

The $d=4$ pole in $S_{1{\rm-loop}}$ can therefore be absorbed by
the following renormalization of the coupling constants
\begin{eqnarray}
\label{renorm}
\begin{array}{ccc}
L_0=L_0^r+\frac{N_f-1}{48} \lambda & L_1=L_1^r+\frac{1}{32}
\lambda & L_2=L_2^r+\frac{1}{16} \lambda \\
 L_3=L_3^r+\frac{2 N_f+1}{48} \lambda &
L_4=L_4^r+\frac{1}{16} \lambda & L_5=L_5^r+\frac{N_f}{8} \lambda \\
L_6=L_6^r+\frac{N_f^2+1}{32 N_f^2} \lambda & L_7=L_7^r &
L_8=L_8^r+\frac{(N_f+1)(N_f-2)}{16 N_f} \lambda \\
&H_2=H_2^r+\frac{(N_f+1)(N_f-2)}{8 N_f}\lambda \ , &
\end{array}
\end{eqnarray}
where 
\be
\lambda= -\frac{\Gamma(-\frac d2)}{(4\pi)^{d/2}}
\Lambda ^{d-4}  
+\frac 1{64\pi^2} \Lambda^{d-4}.
\label{lambda}
\ee
The finite counter terms in $\lambda$ are introduced to partially
cancel 
the finite contributions of the one-loop integrals. This is the
subtraction scheme traditionally used in chiral perturbation theory
\cite{GaL}.  
The cancellation of the finite counterterms
will be discussed in detail below.
In terms of the renormalized coupling
constants $L_i^r$ and $H_2^r$, the sum $S_{1{\rm -loop}}+S_4$
remains finite for  $d\to 4$ both in the normal phase 
and in superfluid phase.
%The coupling constants $L_i^r$ depend on the renormalization scale
%$\Lambda$. One can also define scale independent low-energy 
%coupling constants $\bar{L}_i$:
Notice that the renormalized
 low-energy constants depend on the renormalization scale $\Lambda$.
%\begin{eqnarray}
%\label{renormScaleInd}
%\begin{array}{ccc}
%L_0^r=\bar L_0+\frac{N_f-1}{768\pi^2} \log \frac M\Lambda 
%& L_1^r=\bar L_1+\frac{1}{512 \pi^2} \log \frac M\Lambda 
%& L_2^r=\bar L_2+\frac{1}{256 \pi^2} \log \frac M\Lambda \\
% L_3^r=\bar L_3+\frac{2 N_f+1}{768\pi^2} \log \frac M\Lambda
%&L_4^r=\bar L_4+\frac{1}{256\pi^2} \log \frac M\Lambda 
%& L_5^r=\bar L_5+\frac{N_f}{128 \pi^2} \log \frac M\Lambda   \\
%L_6^r=\bar L_6+\frac{N_f^2+1}{512 \pi^2 N_f^2} \log \frac M \Lambda
%& L_7^r=\bar L_7 &
%L_8^r=\bar L_8+\frac{(N_f+1)(N_f-2)}{256 \pi^2 N_f} 
% \log \frac M\Lambda    \\
%&H_2^r=\bar H_2+\frac{(N_f+1)(N_f-2)}{128 \pi^2 N_f} 
%\log \frac M\Lambda \ . &
%\end{array}  \nonumber \\
%\end{eqnarray}
%
The numerical values of these coupling constants are
not yet known, but they
can be obtained, at least in principle,  from lattice QCD simulations.
For   QCD with {\em three} colors and quarks in the fundamental
representation,
the experimental result for the physical value of these coupling
constants, i.e. $L_i^r$ at $\Lambda=m_\pi$,  
is that they are of the order of $10^{-3}$.  
We expect that this is 
a reasonable estimate for their value for QCD with two colors
and quarks in the fundamental representation as well.

\subsubsection{Wess-Zumino Term}

To account for the chiral anomaly the ${\cal O}(p^4)$ Lagrangian must
be  supplemented by a Wess-Zumino term. At nonzero baryon
density this term can be obtained by gauging the Wess-Zumino term
at zero baryon density \cite{DRS,LSS}. The latter is given by
\be
\Gamma_{WZ}[\Sigma] = \frac {i}{120\pi^2} \int_{M^5} {\rm Tr} \alpha^5,
\ee
where
\begin{equation}
\alpha =\left( \d_\nu \Sigma\right) \Sigma^{-1} dx^\nu  \ ,
\end{equation}
and
$M^5$ is a five dimensional domain with  space-time, $M^4$,
as boundary. 

At non-zero chemical potential the vector field
$B_\nu =\delta_{\nu,0}B$ has only one nonzero component and all
terms in the gauged Wess-Zumino term with more than one external
vector field vanish. We thus find that
\begin{equation}
\Gamma_{WZ}[\Sigma,B_\nu=\delta_{\nu,0}B] = \mu \frac i{12\pi^2}
\int_{M^{4}}{\rm Tr}\left[ B\alpha ^{3}\right]=\mu \frac i{12\pi^2}\,
\int_{M^4}\Tr\left[B\alpha_{i}\alpha_{j}\alpha_{k}\right]
\epsilon^{0ijk} d^4\,x \ .
\nn \\
\label{UniqueLT}
\end{equation}
This term is proportional to $2i$ times the winding number of the
field. The factor 2 arise because $\mu$ is the quark chemical potential
equal to half the baryon number chemical potential for QCD with
two colors. The lightest excitation with nonzero winding
number is expected to be of the order of the mass of the
nucleon for QCD with three colors.
Such terms are subleading in the low-energy limit we are considering
and can be ignored.

\section{Free Energy}

In the previous section we have shown that 
the 1-loop contribution 
to the free energy given by to the logarithm of the determinant
of the propagator matrix is renormalizable. In this section, we obtain
an explicit expression for the  free energy to one-loop order. The
Feynman graphs that enter the free energy at this order are given in
fig.~1. 

\begin{center}
\begin{figure}[ht!]
\vspace{0.5cm}
\hspace*{1cm}
\epsfig{file=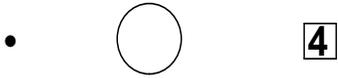, width=4.5cm, height=1cm}
\vspace{-0.5cm}
\caption[]{\small Feynman diagrams that enter into the free energy at
  next-to-leading order. The dot denotes the contribution from ${\cal
    L}^{(2)}$, and the boxed $4$ the contribution from  ${\cal L}^{(4)}$.}
\end{figure}
\end{center}

\subsection{Free Energy of the Normal Phase}

In the normal phase the
one-loop free energy is readily found to be given by
\begin{eqnarray}
  \label{FrEnN}
  \Omega&=& -2  N_f F^2 M^2 - \frac12 (N_f^2-1) \Delta_0^P
-\frac14 N_f (N_f-1) \Delta_0^Q \\
&&-2 N_f M^4  \Big( 8 N_f L_6+2 L_8 +H_2 \Big),  \nonumber
\end{eqnarray}
where $M^2=mG/F^2$ is the leading order pion mass. 
The terms $\Delta_0^{P}$ and $\Delta_0^{Q}$ 
represent the one-loop graphs with a
$P$-mode 
or both $Q$-modes respectively. As we shall see
shortly,
the coupling constants $L_i$ derived in previous section (\ref{renorm})
contain the counter-terms 
that regularize 
these divergent one-loop integrals. 

The one-loop diagrams with a
$P$-mode, or both $Q$-modes are respectively given by (\ref{invPropN})
\begin{eqnarray}
  \Delta_0^P&=&-\int \frac{d^dp}{(2 \pi)^d} \ln (p^2+M^2) 
=\frac1{(4 \pi)^{d/2}} \Gamma(-\frac d2) M^d \\
  \Delta_0^Q&=&-\int \frac{d^dp}{(2 \pi)^d} \ln \Big( (p^2+M^2-4
  \mu^2)^2 
     +16 \mu^2 p_0^2 \Big)
=\frac2{(4 \pi)^{d/2}} \Gamma(-\frac d2) M^d.
\end{eqnarray}
For the $Q$-modes, we simplified the integral using that
\begin{eqnarray}
  (p^2+M^2-4 \mu^2)^2 +16 \mu^2 p_0^2
=\Big( (p_0-2 i\mu)^2+\vec{p}^2+M^2 \Big) \Big((p_0+2 i\mu)^2
+\vec{p}^2+M^2 \Big). \nonumber \\
\end{eqnarray}
Notice that the divergent part of the one-loop diagrams is exactly
canceled by the renormalized coupling constants (\ref{renorm}).
The free energy of the normal phase  reads
\begin{eqnarray}
  \label{FrEnNfinal}
  \Omega = -2 N_f F^2 M^2 \left (1+\frac{M^2}{F^2}\left [ 
8 N_f L_6^r+2 L_8^r+H_2^r 
+\frac {2N_f^2-N_f-1}{256\pi^2N_f} (1-2 \ln \frac{M^2}{\Lambda^2} )
\right ] \right ).\nn \\
\end{eqnarray}
As expected at zero temperature, the free energy of the normal phase 
does not depend on the chemical potential. Therefore the baryon
density is zero in the normal phase. The free energy does not depend
on the renormalization scale $\Lambda$: The logarithmic dependence of
the renormalized coupling constants in $\Lambda$ cancels 
the explicit logarithm in (\ref{FrEnNfinal}). 
Using the free energy (\ref{FrEnNfinal})
the chiral condensate in the normal phase
at next-to-leading order in chiral perturbation theory is given by
\begin{eqnarray}
  \label{qbarq-normalP}
\langle \bar \psi \psi \rangle_0= - \frac{\d \Omega}{\d m} = 2 N_f G
\left (1+2\frac{M^2}{F^2}\left [ 
8 N_f L_6 ^r+2 L_8^r +H_2^r 
-\frac {2N_f^2-N_f-1}{128\pi^2N_f}  \ln \frac{M^2}{\Lambda^2} \right ]
\right ). \nn \\
\end{eqnarray}

%%%%%%%%%%%%%%%%%%%%%%%%%%%%%%%%%%
\subsection{Free Energy of the Phase of Condensed Diquarks}

In the phase of condensed diquarks, the 
free energy is found to be given by
\begin{eqnarray}
  \label{FrEnD}
  \Omega&=& -2 N_f F^2 \Big(\frac12(M_1^2+M_2^2)+\frac 14M_3^2
  \Big)\nn \\ 
&&-\frac14 N_f (N_f+1) \Delta_0^{P_S}-\frac14 (N_f(N_f-1)-2)
\Delta_0^{P_A}-\frac14 N_f (N_f-1) \Delta_0^Q  \nonumber\\
&& -2 N_f (M_1^2-M_2^2)^2 \Big(L_0+2 N_f L_1+2 N_f L_2+L_3 \Big) 
\\
&&-4 N_f (M_1^2-M_2^2)(M_2^2 +\frac 14 M_3^2) \Big(2 N_f
L_4+L_5 \Big) \nonumber \\
&& -8 N_F (M_2^2 +\frac14 M_3^2 )^2 \Big(2 N_f L_6+L_8
\Big)-2
N_f \tilde M^4 \Big(-2 L_8+H_2 \Big) \ , \nn
\end{eqnarray}
where $\tilde M^2=G \sqrt{m^2+j^2}/F^2$ and $\tan \phi=j/m$.

The one-loop diagrams that contribute to the free energy contain a
$P_S$,  a $P_A$ or the mixed $Q$ modes. Their contributions
are given by (\ref{invPropD}) 
\begin{eqnarray}
\label{propag}
  \Delta_0^{P_S}&=&-\int \frac{d^dp}{(2 \pi)^d}
  \ln \Big(p^2+ M_1^2+\frac 14 M_3^2 \Big)
  = \frac{1}{(4 \pi)^{d/2}} \Gamma(-\frac d2)
\Big(M_1^2+\frac 14 M_3^2 \Big)^{d/2} \nn\\
\Delta_0^{P_A}&=&-\int \frac{d^dp}{(2 \pi)^d}
               \ln \Big(p^2+M_2^2 +\frac 14 M_3^2) \Big)
=\frac{1}{(4 \pi)^{d/2}} \Gamma (-\frac d2) \Big(M_2^2
+\frac 14 M_3^2 
\Big)^{d/2} \nonumber\\
 \Delta_0^Q&=&-\int \frac{d^dp}{(2 \pi)^d}
               \ln \Big((p^2+M_1^2)(p^2+M_2^2) +p_0^2 M_3^2) \Big)
\nn \\
&= & 2\frac {\Gamma(-\frac d2)}{(4\pi)^{d/2}}
\Big[\frac 12(M_1^2 +M_2^2) +\frac 14 M_3^2 \Big]^{d/2}
+\frac {\Gamma(-\frac d2)M^{d-4}}{(4\pi)^{d/2}}\frac d8
(M_1^2-M_2^2)^2 
\nn \\
&&+ \frac {(M_1^2-M_2^2)^2}{32\pi^2} \left [\frac 12
-\ln\frac{\sqrt{M_1^2+M_2^2}+\sqrt{M_1^2+M_2^2+M_3^2}}{2M} \right 
. \nn \\ 
&&\left . +\frac{M_1^2+M_2^2}{M_3^2} 
            -\frac{\sqrt{M_1^2+M_2^2}\sqrt{M_1^2+M_2^2 +M_3^2}}{M_3^2}
\right ]
\nn \\ &&+ O((M_1^2-M_2^2)^4)+O((M_1^4-M_2^4)^2) +O(d-4),
\label{deltas}
\end{eqnarray}
respectively. The masses
$M_{1,2,3}$ have been defined in (\ref{massQ}). A derivation of the
power series expansion in $M_1^2-M_2^2$ of
$\Delta_0^Q$ is given in appendix B. The divergent term can be
computed exactly, but we succeeded only to compute the finite corrections near
the leading-order saddle point for $j\ll m$.
One easily shows that the sum of the divergent one-loop terms given by
\be
-\frac{\Gamma(-\frac d2)}{4(4\pi)^{d/2}} \Bigl [N_f^2(
2M_1^4 + 2M_2^4 + \frac 14 M_3^4 +M_3^2(M_1^2+M_2^2))
-2(N_f+1)(M_2^2 + \frac 14 M_3^2)^2 \Bigr ],\nn
\ee
is exactly canceled by the divergent contributions of the counterterms.

In addition to the finite terms already given in the expression for
$\Delta_0^Q$, there are several more finite 
terms that arise from the $d\to 4$ limit of the one-loop diagrams.
The first type of contributions we consider are of the form
\be
\Delta_{I} \equiv \frac{\Gamma(-\frac d2)}{(4\pi)^{d/2}} 
M^{d-4}(1+ S)^{d/2},
\ee
and are present in $\Delta_0^{P_s}$, $\Delta_0^{P_A}$ and
$\Delta_0^Q$.  
To extract the finite terms contributing to $\Delta_I$  we 
express $\Delta_I$ as
\be
\Delta_I 
&=&\left (\frac{\Gamma(-\frac d2)}{(4\pi)^{d/2}} -\frac 1{64\pi^2}
\right) M^{d-4}(1+ S)^{2} 
+\frac 1{32\pi^2} M^{d-4} (1+S)^2(\frac 12 -\ln(1+S))
\nn \\ &&+O(d-4)\nn \\
&=&-\lambda \left( \frac M \Lambda \right)^{d-4} (1+S)^2+ \frac1{64
  \pi^2} M^{d-4} (1-S^2)^2+O(S^3)+O(d-4). 
\label{fp4}
\ee
%In our subtraction scheme, with "$-1/64\pi^2$" as additional 
%finite counter term,
%the divergent terms in the  
%combination 
%\be
%\left (\frac{\Gamma(-\frac d2)}{(4\pi)^{d/2}} -\frac 1{64\pi^2}
%\right)\left(\frac M\Lambda \right)^{d-4}(1+ S)^{2}
%\ee
% is absorbed by the  next-to-leading order coupling constants. One
%easily shows that 
%finite terms linear in $S$ are absent. 
%To order $S^2$,
%the renormalized value of $\Delta$ is thus given by
%\be
%\Delta_R = \frac{(1-S^2)^2}{64\pi^2} +O(S^3).
%\label{fp4}
%\ee
The second type of one-loop contributions we have to consider in the
limit $d \to 4$ are of the form
\be
\Delta_{II} \equiv 
\frac{\Gamma(-\frac d2)}{(4\pi)^{d/2}} \frac d8 
M^{d-4}(M_1^2-M_2^2)^2.
\ee
It occur only in $\Delta_0^Q$.
By expanding in powers of $d-4$ we find that 
\be
\Delta_{II} 
&=& \left [ \frac{\Gamma(-\frac d2)}{(4\pi)^{d/2}} -\frac 1{64\pi^2}
\right ] \frac 12  M^{d-4}(M_1^2-M_2^2)^2  +
O(d-4) \nn \\
&=&-\frac12 \lambda \left( \frac M \Lambda \right)^{d-4}
(M_1^2-M_2^2)^2+O(d-4). 
\ee
Thus, in our subtraction scheme, there are no finite terms resulting
from such contributions.

Using (\ref{fp4}) one easily finds the following finite contributions
from the one-loop diagrams to second order in $M_1^2 -M_2^2$ and
$M_2^2+M_3^2/4-M^2$,
\be
&-&\frac {2N_f^2-N_f-1}{128 \pi^2}  
+\frac {N_f^2+N_f}{128 \pi^2} 
(M_1^2-M_2^2)^2  
+\frac {2N_f^2-N_f-1}{64 \pi^2} (M_2^2 +\frac 14 M_3^2 -M^2)^2 \nn \\
&+&\frac {N_f^2}{32 \pi^2}(M_1^2-M_2^2)(M_2^2 +\frac 14 M_3^2 -M^2).
\ee
The second term also includes the finite contribution to $\Delta_0^Q$
given in (\ref{deltas}).

%%%%%%%%%%%%%%%%%%%%%%%%%%%%%%%%

\section{Landau Ginzburg Model}

In this section we study the phase transition that separates
 the normal phase from the phase of condensed diquarks.
At the mean field level and for zero diquark source,
it was found that there is a second order
 phase transition at $\mu=M/2$, where $M$ is the pion mass to leading
 order in chiral perturbation theory.
 The order parameter of this phase transition is the diquark
 condensate. To leading order it was found that $\langle \psi \psi
 \rangle=2 N_f G \sin \alpha$, with $\alpha=\arccos (M^2/4 \mu^2)$
 \cite{KSTVZ}.

Because of the contributions of the $Q$-integrals,  the 
next-to-leading order analytical expressions for the free energy 
are quite complicated. 
However, the neighborhood of the critical point
 can be easily studied by means of a Landau-Ginzburg theory for the 
order parameter. 
~From the leading order result we find that $\alpha$ is a suitable order
 parameter 
near the critical point. This is quite natural since $\alpha$ is the
rotation angle of the condensate. Therefore if we expand the free
energy in 
powers of $\alpha$ for $\alpha\ll 1$, we should obtain
a Landau-Ginzburg model describing the phase transition. Notice,
however, that this Landau-Ginzburg model results from an exact
calculation within the effective theory. Furthermore in this approach
$\alpha$ has to be an independent variable so that
the saddle point equation cannot be used to express $\alpha$ in terms
of the 
pion mass and the chemical potential. We first illustrate the
usefulness of 
the expansion of the free energy in powers of $\alpha$ by analyzing
the theory at leading order and then  consider the free
energy at next-to-leading order.

At the mean field level, the free energy in the diquark condensation
phase just above the critical chemical potential, i.e. for
$\alpha\ll1$, is given by
\begin{eqnarray}
  \label{LGlo}
  \Omega=-N_f F^2 \Big( 2 M^2+2 M^2\phi \alpha
+(4 \mu^2-M^2) \alpha^2 -\frac1{12} (16 \mu^2-M^2) \alpha^4 \Big)
+O(\alpha^6),
\end{eqnarray}
where the diquark source enters through $\phi=\tan(j/m)\simeq j/m$ 
for $j\ll m$.
This free energy can be studied as a Landau-Ginzburg model.
At $\phi=0$ there is a second-order phase transition where
the coefficient of the $\alpha^2$-term vanishes,
that is at $\mu=M/2$. For zero diquark source,
the order parameter obtained from the minimum
of the effective potential is given by
\begin{eqnarray}
  \label{OPlo}
  \alpha=\sqrt{\frac{6 (4 \mu^2-M^2)}{16 \mu^2-M^2}}.
\end{eqnarray}
At the critical point for nonzero diquark source we find the usual
mean 
field value of the critical exponent with $\alpha$ given by
\begin{eqnarray}
  \label{CIlo}
  \alpha^3=2 \phi,
\end{eqnarray}
Notice that
this relation implies that $\phi \ll \alpha \ll 1$.
Therefore the diquark condensate near the critical point is given by
\begin{eqnarray}
  \label{DQlo}
  \langle \psi \psi \rangle=-\frac{\partial \Omega}{\partial j}
  \Big|_{j=0} 
=-\frac1m \frac{\partial \Omega}{\partial \phi}
\Big|_{\phi=0}=2 N_f G \alpha
\end{eqnarray}
where we have used that to leading order $mG=M^2 F^2$. 
To leading order in a $(\mu-M/2)$ expansion this result agrees
with the exact mean field theory result.

At next-to-leading order the situation is more complicated. 
We study the free energy for a chemical potential near the
leading-order 
critical point and define $\bar \mu$ by $\mu=M/2+M \bar \mu$.
The leading order results provides us with power counting rules for
$\alpha$, 
$\bar \mu$ and $\phi$ near the critical chemical potential: 
if $\alpha\sim \epsilon$, then $\bar \mu\sim
\epsilon^2$ and $\phi \sim \epsilon^3$. 
%We need the diquark source to
%first order only, therefore we will restrict ourselves to first order
%in $\phi$. 
To obtain an expansion to fourth order in $\epsilon$, we use
\be
\frac{M_1^2-M_2^2}{M^2} &=& \frac{4\mu^2\sin^2\
\alpha}{M^2} =
\alpha^2 -\frac 13 \alpha^4+4\alpha^2\bar \mu +\cdots,\nn\\
\frac{M_2^2+\frac 14 M_3^2}{M^2} - 1 &=& \cos(\alpha-\phi)-1=
-\frac12 \alpha^2 +\frac 1{24}\alpha^4
+ \phi\alpha +\cdots.
\nn\\ 
\ee
To this order the free energy  is given by
\begin{eqnarray}
  \label{LGnlo}
  \frac{\Omega}{-2 N_f F^2 M^2}&=&
1+ a_0 \frac {M^2}{F^2} + \left(1+ a_1\frac{M^2}{F^2} \right )
  \phi \alpha + a_2\frac{M^2}{F^2} \alpha^2
\nn \\  
&&+ \left (2 +  a_3\frac{M^2}{F^2} \right )\bar \mu \alpha^2 
+ \left(-\frac18+ a_4\frac{M^2}{F^2} \right ) \alpha^4,
\ee
where the coefficients of the next to leading order corrections are
  given by
\be
a_0 &=&   8N_f L^r_6 +2 L^r_8 +H^r_2  
+ \frac {2N_f^2-N_f-1}{256 \pi^2 N_f} (1-2  \ln
\frac{M^2}{\Lambda^2}),\nn \\
a_1 &= &   8 (2N_f L^r_6+L^r_8) -\frac{2N_f^2-N_f-1}{64
      \pi^2 N_f}  \ln \frac{M^2}{\Lambda^2}  ,\nn \\
a_2 &=&  2(2N_f L^r_4+L^r_5-4N_f
    L^r_6-2L^r_8) -\frac{N_f+1}{128
      \pi^2 N_f}  \ln \frac{M^2}{\Lambda^2}, \nn \\
a_3 &=& 8(2N_fL^r_4+L^r_5) -\frac{N_f}{16
      \pi^2 }  \ln \frac{M^2}{\Lambda^2} , \nn \\
a_4 &=&   -\frac{5}{3}(2N_fL^r_4+L^r_5)  
+   \frac 43(2N_fL^r_6+L^r_8) 
 +(L^r_0+2N_fL^r_1+2N_f L^r_2+L^r_3)   \nn \\
&& -\frac {N_f-1}{512 \pi^2 N_f} +\frac{N_f+1}{384
      \pi^2 N_f}  \ln \frac{M^2}{\Lambda^2}. 
\nn \\
\ee
This free energy is typical of a system exhibiting a 
second order phase transition.

For zero diquark source, the phase transition occurs when the
coefficient of the $\alpha^2$-term in the free energy vanishes.
Therefore we find that there is a second order phase transition at
\begin{eqnarray}
\label{CRnlo}
\bar \mu_c &=& -\frac 12  a_2 \frac{M^2}{F^2}\nn \\
&=&\frac{M^2}{F^2}\left[ -2N_f
 L^r_4- L^r_5+4 N_f L^r_6+2 L^r_8+\frac{N_f+1}{256
      \pi^2 N_f}  \ln \frac{M^2}{\Lambda^2} \right] +
  O(\frac{M^4}{F^4}). 
\end{eqnarray}  
We compare this result to the pion mass at zero chemical potential
at next to leading order in
chiral perturbation theory (see Appendix~A) ,
\be
m_\pi^2=M^2 \left( 1+\frac{M^2}{F^2} 
 \left[ 4(-2N_f L^r_4- L^r_5+4N_f L^r_6+2 L^r_8)+\frac{N_f+1}
   {64  \pi^2 N_f}  \ln \frac{M^2}{\Lambda^2} \right] \right).
\ee
Therefore, the critical chemical potential is given by 
\begin{eqnarray}
\mu_c=M/2+M \bar \mu_c=m_\pi/2.
\end{eqnarray}

For a nonzero diquark source we find that at the critical point
\begin{eqnarray}
\label{crit}
\alpha^3&=&2\phi\left ( 1 +[a_1+8a_4] \frac {M^2}{F^2} \right )\nn \\
&=&                     
2\phi\left(1+\frac{M^2}{F^2} \left[  -\frac{40}3 (2
    N_f L^r_4+L^r_5) +\frac{56}3 (2N_f L^r_6+L^r_8) \right .\right .  \\
&&+ \left . \left . 8
    (L^r_0+2N_fL^r_1+2N_f L^r_2+L^r_3) 
-\frac {N_f -1}{64 \pi^2 N_f} -\frac{6 N_f^2-7N_f-7}{192 
      \pi^2 N_f}  \ln \frac{M^2}{\Lambda^2}\right] \right). \nn
\end{eqnarray} 
There are no logarithmic corrections in $\phi$ or $\alpha$ so that the
critical exponent remains at its mean field value. 
This is somewhat surprising. For zero diquark source, half of the
$Q-$modes are true massless Goldstone modes. They result from the
breaking of the remaining symmetries of the microscopic theory at
nonzero chemical potential and nonzero quark mass 
by the diquark condensate \cite{KSTVZ}. 
For nonzero diquark source, these modes become
massive. They are pseudo-Goldstone modes, 
with a square mass equal to $M^2 \phi (3 \alpha+8 \phi)/32$ at
leading order at $\mu=M/2$ (\ref{invPropD}). 
Usually loop-graphs with 
pseudo-Goldstone modes produce logarithmic terms of the type found 
for instance in 
the free energy of the normal phase (\ref{FrEnNfinal}). In the
superfluid 
phase, we would have therefore expected one-loop corrections to the
free energy of the
form $\alpha \phi \ln(\phi)$. These terms would produce a
deviation from 
mean-field critical indices in (\ref{crit}). However, we do not find
any logarithmic term neither in $\phi$ nor in $\alpha$.
This is a non-trivial result, and one should
expect that to one-loop order the critical exponents of the
microscopic theory 
are given by mean field theory.

For zero diquark source, the value of the order parameter 
$\alpha$ is given by the minimum of the free energy. It reads 
\begin{eqnarray}
  \label{alpha}
  \alpha^2&=& 4 \left(2 \bar \mu+\frac{M^2}{F^2}(a_2+a_3 \bar \mu)
  \right) \left(1+8 
  \frac{M^2}{F^2} a_4 \right) \nn\\
&=&\frac{8}{m_\pi} \left(\mu-\frac{m_\pi}2 \right) 
\Bigg(1+\frac{m_\pi^2}{F_\pi^2} \left[ -\frac{34}3 (2 N_FL^r_4+L^r_5)
  + \frac{44}3 (2N_f L^r_6+L^r_8) \right. \\
&&\left . +8 (L^r_0+2N_fL^r_1+2N_f L^r_2+L^r_3)
-\frac {N_f-1}{64 \pi^2 N_f} -\frac{12 N_f^2-11N_f-11}{384
      \pi^2 N_f}  \ln \frac{M^2}{\Lambda^2 }\right]  \Bigg) \nn
\end{eqnarray}
Therefore we find that the order parameter $\alpha$ vanishes for
$\mu \leq m_\pi/2$ and that it increases continuously for higher
chemical potential. The phase transition is thus second order.

\section{Number Density and Condensates}

The phase structure of QCD with two colors at nonzero baryon chemical
potential can be characterized by three important observables:
the quark number density, the
chiral condensate and the diquark condensate. 
We compute them at zero diquark source.
The number density is obtained from the free energy as follows 
\begin{eqnarray}
n_B&=&-\frac{d \Omega}{d \mu}=-\frac1M \frac{\partial \Omega}{\partial
  \bar \mu} \nn \\
&=& 4N_f M F^2\left(1+\frac12 a_3  \frac {M^2}{F^2}\right) \alpha^2\nn \\
&=&
4 N_f M F^2 \alpha^2 \left( 1 +\frac{M^2}{F^2}\left[
    4(2N_fL^r_4+L^r_5) -\frac{N_f}{32
      \pi^2}  \ln \frac{M^2}{\Lambda^2}\right] \right)\nn
\\
&=&4 N_f M^2 F_\pi^2 \alpha^2.
\end{eqnarray}

The chiral condensate is given by
\begin{eqnarray}
  \langle \bar \psi \psi \rangle&=&-\frac{d \Omega}{d m}
=-\frac{G}{F^2} \left( \frac{\partial \Omega}{\partial M^2}+\frac{d
    \bar 
  \mu}{d M^2} \frac{\partial \Omega}{\partial \bar \mu}
\right)=-\frac{G}{F^2} \left( \frac{\partial \Omega}{\partial
    M^2}-\frac{1+2 \bar \mu}{4 M^2}
\frac{\partial \Omega}{\partial \bar \mu} \right) \nn \\
&=&2 N_f G \left( 1-\frac 12 \alpha^2
+\frac{M^2}{F^2}\left[2 a_0 -\frac{2N_f^2-N_f-1}{128 \pi^2 N_f} +
  \left(2a_2-\frac 14 a_3 
-\frac{N_f+1}{128\pi^2 N_f}\right)\alpha^2 \right ]\right )\nn \\
&=&\langle \bar \psi \psi \rangle_0 \left( 1- \frac12 \alpha^2
    +\frac{M^2}{F^2} \Bigg[
 4 N_f L^r_4+2 L^r_5-8 N_f L^r_6-6 L^r_8 +H_2^r
\right. \\ 
&&\left.  \hspace{5cm}  -\frac{N_f+1}
{128\pi^2 N_f}  -\frac{N_f+1}{128\pi^2 N_f}  
\ln \frac{M^2}{\Lambda^2} \Bigg] 
\alpha^2 \right). \nn
\end{eqnarray}

Finally the diquark condensate is given by
\begin{eqnarray}
\langle \psi \psi \rangle&=&-\frac{d \Omega}{d j}\Big|_{j=0}= -\frac1m
\frac{d \Omega}{d \phi}\Big|_{\phi=0} \nn \\
&=& 2N_f G \left(1+a_1  \frac {M^2}{F^2} \right)\alpha\nn \\
&=&
2 N_f G   \left( 1+ \frac{M^2}{F^2} \left[8  (2N_f
    L^r_6+L^r_8)-\frac{2N_f^2-N_f-1}{64  
      \pi^2 N_f}  \ln \frac{M^2}{\Lambda^2} 
    \right]    \right) \alpha \nn \\
&=&\langle \bar \psi \psi \rangle_0 \left( 1+ \frac{M^2}{F^2} 2
    (2 L^r_8-H^r_2)  \right) \alpha.
\end{eqnarray}

The baryon number density and the diquark condensate vanish at
$\mu=m_\pi/2$. They increase continuously for $\mu \geq m_\pi/2$, 
whereas the quark-antiquark condensate diminishes. 
At leading order it was found that in the superfluid phase 
$\langle \psi \psi \rangle^2+\langle \bar \psi \psi \rangle^2 =
\langle \bar \psi \psi \rangle^2_0$, where $\langle \bar \psi \psi
\rangle_0$ is the quark-antiquark condensate in the normal phase. 
This relation does not hold at next-to-leading order. Other than that,
the qualitative behavior of these observables is the same at leading
order and at next-to-leading order.

\section{Conclusions}

As is the case for QCD with three colors, the low-energy sector of 
QCD with two colors and quarks in the fundamental representation 
is a theory of weakly interacting Goldstone bosons. What distinguishes
QCD with two colors from QCD with three colors is that  we
have both mesonic and baryonic Goldstone bosons (also known as
diquarks). 
Therefore, the phase transition to a phase of condensed diquarks can 
be described entirely by means of a low-energy effective theory which
is completely determined by the symmetries of the microscopic theory.
In earlier work, in which this theory was analyzed at the mean field
level, 
it was found that such phase transition takes place at a baryonic
chemical potential equal to half the mass of the Goldstone bosons. 
Above this transition point the chiral condensates rotates into
a diquark condensate with a rotation angle that is determined by
the chemical potential.
Meanwhile, this result has been observed in lattice QCD simulations.

In this article we have analyzed the one-loop corrections of the
leading 
order effective Lagrangian
as well as the tree-graph contributions of
the next-to-leading order terms in the effective Lagrangian.
We have shown that the theory is renormalizable, i.e. all infinities
generated 
by the one-loop diagrams can be absorbed into a 
background field independent redefinition of
 the coupling constants of the next-to-leading order terms in the
effective Lagrangian.  We have derived the counter-terms for a general
background field, and by an explicit calculation, we have verified that 
they cancel the one-loop divergences both in the normal phase and
in the phase of condensed diquarks. 

The one-loop corrections of the effective theory do not qualitatively
change the 
predictions obtained from the  mean field analysis. In particular we 
have obtained the physically 
satisfying result that a second order phase transition to a phase of
condensed diquarks 
takes place at a chemical potential equal
to half the one-loop renormalized pion mass. From the effective theory,
we have derived a Landau-Ginsburg theory for this phase transition 
with the rotation angle of the  chiral condensate as order parameter.
The next-to-leading order corrections do not qualitatively change the
mean field  coefficients. 
The normal phase is characterized by a nonzero quark-antiquark
condensate, a zero diquark condensate and a zero baryon number
density. A second order phase transition at $\mu=m_\pi/2$ separate the
normal phase from the diquark superfluid phase, where the
quark-antiquark condensate, the diquark condensate and the baryon
number density are nonzero.
However, they do affect the magnitude of
chiral condensate and the diquark condensate in a different way so that
they are no longer related by the tangent of the rotation angle.

At the phase
transition point, half of the baryonic mesons become massless. 
The Goldstone bosons associated with the formation of a diquark
condensate remain massless above the phase transition point. 
At the one-loop level these massless modes do not
produce any infrared singularities. Neither do we find any
chiral logarithms related the modes that become massive
above the critical chemical potential. A nonzero diquark
source does not lead to  chiral logarithms either. 
These results corroborate with the absence
of one-loop corrections to the mean field critical exponents. 
The only nonanalytic behavior in the free energy that 
we find is of the form
$\alpha^4 \sqrt{\alpha^2-2\bar\mu}$ (with $\alpha$ the rotation angle
of the chiral condensate and $\bar \mu$ the deviation from the
critical chemical potential). This statement should also hold at the
two-loop level. The analytic structure of the propagator near the 
critical point suggests that critical exponents remain at their 
mean field value to 
all (finite) orders in chiral perturbation theory.

Our results have been derived for QCD with two colors and quark in the
fundamental representation. 
Very similar low-energy effective theories
have been derived  for QCD
with quarks in the adjoint representation and 
for QCD with three or more colors and fundamental quarks but
with the fermion determinant replaced
by its absolute value. The latter theory can be interpreted as QCD
at nonzero isospin chemical potential 
and can be generalized
to a theory with a chemical potential for each quark flavor again
resulting in a similar low-energy effective theory. A closely related
effective theory has been derived in condensed matter physics 
for disordered systems with an imaginary vector potential.
We expect that our conclusions for QCD with two colors will also
be valid for each of these theories. In particular we expect 
the next-to-leading order corrections do not
qualitatively change the mean field behavior.

All our results have been derived for zero temperature. 
As should be the case for a renormalizable theory,
no additional ultraviolet divergencies appear at nonzero temperature. 
Just above the critical chemical potential we expect a second order
phase transition to the normal phase at a critical temperature that 
is in the domain of validity of our chiral Lagrangian. We thus
are able to obtain exact results for the critical temperature within
the framework of chiral perturbation theory. These results as well
as other properties of the diquark phase at nonzero temperature will
be discussed in a forthcoming publication.

{\bf Acknowledgements:} 
G. Baym, J. Kogut, T. Sch\"afer, and J. Skullerud  are acknowledged
for useful discussions. D.T. is supported in part by
``Holderbank''-Stiftung. This work was partially supported by the US
DOE grant DE-FG-88ER40388.

\vskip 1.5cm
\noindent 

\renewcommand{\theequation}{A.\arabic{equation}}
\noindent
{\large \bf Appendix A: One-loop expressions for {\boldmath $m_\pi$}
  and {\boldmath $F_\pi$}}

In QCD with three colors and quarks in the fundamental representation,
the usual way to compute the pion mass and the pion decay constant is
to extract them from the axial two-point correlation function. 

For the pion mass, the Feynman diagrams of fig.~2 have to be
evaluated:.
\begin{center}
\begin{figure}[ht!]
\vspace{0.5cm}
\hspace*{1cm}
\epsfig{file=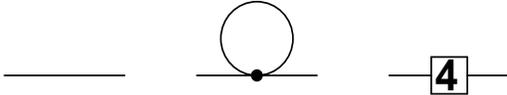, width=6.75cm, height=1.5cm}
\vspace{-0.5cm}
\caption[]{\small Feynman diagrams that enter into the axial-vector
  two-point correlation function that contribute to the pion mass at
  next-to-leading order. The dot denotes a vertex from ${\cal
    L}^{(2)}$, and the boxed $4$ a vertex from  ${\cal L}^{(4)}$.}
\end{figure}
\end{center}

\noindent
The pion mass is given by the pole of this two-point function in
momentum space. However, as will be explained next,
we do not need an explicit calculation of
this two-point function.
The key observation is that the only
finite one-loop contributions originate from the integral in the 
second diagram.
At zero chemical potential, all the Goldstone modes have the same
mass. 
Therefore the one-loop diagram is proportional to
\begin{eqnarray}
\label{tadpole}
\int \frac{d^dp}{(2 \pi)^d} \frac 1{p^2+M^2} &=&
-\frac d2 M^{d-2}  \frac{\Gamma (-\frac d2)}{ (4\pi)^{d/2}}
=2 M^2 \lambda \left (\frac M\Lambda \right )^{d-4}.
\ee
Because our theory is renormalizable, in our subtraction scheme
divergent contributions of
the type given in the last line are cancelled by 
contributions from
the counter terms. The pion mass 
%thus does not receive any contributions
%from one-loop diagrams. It 
is thus completely determined by the tree graphs
of ${\cal L}^{(2)}$ and ${\cal L}^{(4)}$. 

The correction to the pion mass from the tree-graphs 
${\cal L}^{(4)}$ is given by twice the
coefficient of the quadratic term in the pion fields
in ${\cal L}^{(4)}$. It is 
essential to impose the Euclidean on-shell condition $p^2=-M^2$
in the $O(p^4)$ terms of the effective Lagrangian {\cite{GaL}. 
For example,
\be
\Tr[\d_\nu\Sigma\d_\nu\Sigma^\dagger] = - M^2 \pi^a \pi^a + \ldots 
\ee
The on-shell condition results in a contribution from the
$L_4$ and 
$L_5$ terms in ${\cal L}^{(4)}$  eventhough $\mu=0$. Calculating all
traces in ${\cal L}^{(4)}$ to order $\Pi^2$ we find
\be
{\cal L}^{(4)}= \ldots
+\frac{2M^4}{F^2} \left[  -2N_f{L}_4 -{L}_5+
  4N_f {L}_6+2 {L}_8\right] \pi^a \pi^a +\ldots
\ee
The pion mass is therefore given by
\be
m_\pi^2 = M^2\left(1+\frac{4M^2}{F^2} \left[- 2N_f L_4 -L_5+ 
  4 N_f L_6 +2 L_8\right] \right) +{\rm tadpoles}.
\ee
As argued before, our subtraction scheme is such that the tadpoles and
the divergent counter terms in the combination (\ref{lambda}) cancel.
We thus find the renormalized pion mass
\be
m_\pi^2=M^2 \left( 1+\frac{M^2}{F^2} 
 \left[ 4(-2N_f L^r_4- L^r_5+4N_f L^r_6+2 L^r_8)+\frac{N_f+1}
   {64  \pi^2 N_f}  \ln \frac{M^2}{\Lambda^2} \right] \right).
\label{mpi}
\ee
with renormalized coupling constants defined in (\ref{renorm}).

We can proceed in the same way for the pion decay constant. The
contributions from the tree diagrams of ${\cal L}^{(2)}$ and ${\cal
  L}^{(4)}$ are easy to obtain. They are given by the coefficient of
the $\d_\mu \pi^a \d_\mu \pi^a$-term in the effective Lagrangian. The
contribution from the one-loop diagrams are again proportional to
(\ref{tadpole}). We can therefore apply the same method as the one we
used to determine the pion mass. We find that the pion decay constant
at next-to-leading order reads
\be
F_\pi^2 = F^2 \left( 1 +\frac{M^2}{F^2}\left[
    4(2N_fL^r_4+L^r_5) -\frac{N_f}{32
      \pi^2}  \ln \frac{M^2}{\Lambda^2}\right] \right).
\label{Fpi}
\ee

\renewcommand{\theequation}{B.\arabic{equation}}
\noindent
{\large\bf Appendix B: Q-Modes in the Diquark Condensation Phase}

In this appendix we evaluate the one-loop contribution of the $Q$-modes
to the free energy. We thus analyze the following integral
\begin{eqnarray}
 \Delta_0^Q&=&-\int \frac{d^dp}{(2 \pi)^d}
 \ln \Big( (p^2+M_1^2)
(p^2+M_2^2) +p_0^2 M_3^2 \Big), \nonumber
\end{eqnarray}
where $M_1^2$, $M_2^2$, and $M_3^2$ are defined in (\ref{massQ}). 
The pole part of this integral at $d=4$ can be easily obtained
by expanding the logarithm in inverse powers of the momentum,
\be
\ln \Big( (p^2+M_1^2)
(p^2+M_2^2) +p_0^2 M_3^2 \Big) &=& 2\ln p^2 +\ln \Big( 
1+ \frac{M_1^2 +M_2^2}{p^2} +\frac{M_1^2 M_2^2}{p^4} 
+\frac{p_0^2 M_3^2}{p^4} \Big)\nn \\
&=& 2\ln p^2 +
1+ \frac{M_1^2 +M_2^2}{p^2} +\frac{M_1^2 M_2^2}{p^4} 
+\frac{p_0^2 M_3^2}{p^4} \nn \\
&&-\frac 12 \frac{(M_1^2 +M_2^2)^2}{p^4}
-\frac 12 \frac{p_0^4 M_3^4}{p^8}
-\frac{p_0^2M_3^2(M_1^2+M_2^2)}{p^6} \nn \\ &&+ \cdots 
\ee
The higher orders in the expansion in inverse powers of the momentum
vanish in dimensional regularization.
The pole term in $d-4$ is thus given by the integral
\be
 PP(\Delta_0^Q)&=&-\frac{\Omega_{d-1}}{(2\pi)^d}\int_0^\infty p^{d-5}dp
\int_0^\pi d\theta \sin^{d-2}\theta \nn\\
&&\times \left [ 
M_1^2 M_2^2 
-\frac 12 (M_1^2 +M_2^2)^2 -M_3^2(M_1^2+M_2^2)\cos^2\theta
-\frac 12 M_3^4 \cos^4\theta \right ]\nn \\
\label{PP}
\ee
where we have introduced polar coordinates with $p_0 =p\cos\theta$, 
and the area of the $d$-dimensional unit sphere given by
\be
\Omega_{d-1} = \frac{2 \pi^{(d-1)/2}}{\Gamma\left(\frac{d-1}2\right)
  }. 
\ee
The integrals in (\ref{PP}) 
can be easily calculated using the formulae
\begin{eqnarray}
\int_0^\infty dx \frac {x^\beta}{(x^2+M^2)^\alpha} &=& \frac 12
\frac{\Gamma\left (\frac {\beta+1}2\right )
\Gamma\left (\alpha -\frac {\beta+1}2\right )}{\Gamma(\alpha)
(M^2)^{\alpha-(\beta+1)/2}}\\
\int_0^{\pi/2}
d\theta\sin^{\alpha}\theta \cos^\beta \theta&=&
\frac{\Gamma(\frac{1+\alpha}2) 
  \Gamma(\frac{1+\beta}2)}{2 \Gamma(1+\frac{\alpha+\beta}2)}.
\label{TV1}
\end{eqnarray}
For the  $p$-integral one can derive the limiting relation for $d\to 4$,
\be
PP\left(\int_0^\infty p^{d-5}dp\right ) = \Gamma(-\frac d2).
\ee
The final result for the pole part of $\Delta_0^Q$ is thus given by
\be
 PP(\Delta_0^Q) =\frac{\Gamma(-\frac d2)}{(4\pi)^{d/2}} \Bigl ( M_1^4 +M_2^4
+\frac 12(M_1^2+M_2^2)M_3^2 +\frac 18 M_3^4 \Bigr ).
\ee

The finite contributions to $\Delta_0^Q$ can only be obtained close to the
transition point ($\alpha \to 0$, $\mu \to M/2$ and $j \to 0$). 
As starting point the integral is rewritten as
\be
 \Delta_0^Q&=&-\int \frac{d^dp}{(2 \pi)^d}
 \ln \Big( (p^2+\frac{M_1^2+M_2^2}2)^2 
+p_0^2 M_3^2 -\frac {(M_1^2-M_2^2)^2}4\Big). \nonumber
\ee
We wish to compute
the $Q$-mode contribution to the free energy up to $O(\alpha^6)$ and
to first order in the diquark source $j$, i.e. $\phi$. To this order
the integral is given by
\be
 \Delta_0^Q&=&-\int \frac{d^dp}{(2 \pi)^d}
 \ln \Big( (p^2+\frac{M_1^2+M_2^2}2)^2+ 
p_0^2 M_3^2 \Big) \\
&&
 +\int \frac{d^dp}{(2 \pi)^d} \frac{(M_1^2-M_2^2)^2}{4
[(p^2+\frac{M_1^2+M_2^2}2)^2+ 
p_0^2 M_3^2]}, \nonumber
\ee
The first integral can be simplified by shifting the
$p_0$-integration. Introducing $d$-dimensional polar coordinates 
as in the calculation of the pole part we then find
to order $\alpha^6$.
\be
 \Delta_0^Q&=&\frac{{\Omega}_{d-1}}
{(2\pi)^d} \int_0^\infty dp p^{d-1}
\int_0^\pi d\theta \sin^{d-2} \theta \left [ -2\ln (p^2 +
\frac12(M_1^2+M_2^2) +\frac 14M_3^2) \right . \nn \\
&&\left . 
 +\frac{(M_1^2-M_2^2)^2}{4[(p^2+\frac12 (M_1^2+M_2^2))^2 
+p_0^2 M_3^2]} \right ] +O((M_1^2-M_2^2)^4) , \nonumber\\
&=&\frac{{\Omega}_{d-1}}{(2\pi)^d} \int_0^\infty dp
p^{d-1} 
\int_0^\pi d\theta \sin^{d-2} \theta \left [-2 \ln (p^2 +
\frac12(M_1^2+M_2^2) +\frac 14M_3^2) \right . \\
&& \left .  
 +\frac{(M_1^2-M_2^2)^2}{4p^2[p^2 
+M_1^2+M_2^2 +\cos^2\theta M_3^2]} 
\right ] +O((M_1^2-M_2^2)^4)+O((M_1^4-M_2^4)^2)
\nonumber
\ee
The final result is
\be
 \Delta_0^Q&= & 2\frac {\Gamma(-\frac d2)}{(4\pi)^{d/2}}
\Big[\frac 12(M_1^2 +M_2^2) +\frac 14 M_3^2 \Big]^{d/2}
+\frac {\Gamma(-\frac d2)}{(4\pi)^{d/2}}\frac d8 
M^{d-4}(M_1^2-M_2^2)^2
\nn \\
&&+ \frac {(M_1^2-M_2^2)^2}{32\pi^2} \left [\frac 12
-\ln\frac{\sqrt{M_1^2+M_2^2}+\sqrt{M_1^2+M_2^2+M_3^2}}{2M} \right
. \nn \\ 
&&\left .+\frac{M_1^2+M_2^2}{M_3^2} -\frac{\sqrt{M_1^2+M_2^2+M_3^2} 
\sqrt{M_1^2+M_2^2}}{M_3^2}
\right ] \\
&&+O((M_1^2-M_2^2)^4)+O((M_1^4-M_2^4)^2) +O(d-4). \nn 
\ee 
This result correctly reproduces the pole part of $\Delta_0^Q$ which
was obtained at the beginning of this section.

\end{document}